\documentclass[aps,prb,preprintnumbers,reprint,amsmath,amssymb,floatfix,superscriptaddress]{revtex4-1}

\usepackage{graphicx}
\usepackage{bm}
\usepackage{amsmath}
\usepackage{amssymb}
\usepackage{bbold}
\usepackage{braket}
\usepackage{color}
\usepackage{xspace}
\newcommand{\Mo}{Mo$_3$S$_7$(dmit)$_3$\xspace}

\graphicspath{{figures/}}

\begin{document} 

\title{Haldane insulator protected by reflection symmetry in the doped Hubbard model on the three-legged ladder}

\author{H. L. Nourse}
\email{hnourse@gmail.com}
\affiliation{School of Mathematics and Physics, The University of Queensland, Brisbane, Queensland 4072, Australia}
\author{I. P. McCulloch}
\affiliation{Centre for Engineered Quantum Systems, School of Mathematics and Physics, The University of Queensland, Brisbane, Queensland 4072, Australia}
\author{C. Janani}
\affiliation{Centre for Engineered Quantum Systems, School of Mathematics and Physics, The University of Queensland, Brisbane, Queensland 4072, Australia}
\author{B. J. Powell}
\affiliation{School of Mathematics and Physics, The University of Queensland, Brisbane, Queensland 4072, Australia}

\begin{abstract}
We demonstrate the existence of an insulating phase in the three-legged Hubbard ladder at two-thirds filling. In this phase chargons are bound because the physics within a unit cell favors the formation of triplets. The resultant moments lead to a ground state in the Haldane phase, a symmetry protected topological state of matter. In this purely fermionic model, reflection is protecting but time reversal and dihedral symmetries are not, in contrast to spin models.
\end{abstract}

\maketitle 

\section{Introduction}

The Mott--Hubbard metal-insulator transition is one of the central paradigms for strongly correlated electron physics \cite{mott1949,gebhard1997}. In the single band Hubbard model \cite{hubbard1963,hubbard1964,gutzwiller1963,gutzwiller1965,kanamori1963} on-site interactions cause an insulating state due to the binding of charged excitations (chargons).\cite{dallatorrePRL2006,Endres}
%
%
Away from half-filling the single band Hubbard model is believed to be metallic. Here we show that, contrary to this expectation, the Hubbard model on the three--legged ladder has another correlated insulating state at two-thirds--filling. Here chargons are bound by the interplay of on-site interactions with interference effects within the unit cell, which drive triplet formation. We conjecture that similar insulating states occur at  commensurate fillings in other odd--legged ladders. Furthermore, we argue that  this state is realized in \Mo. \cite{llusar2004,jackoPRB2015}

\Mo displays a large charge gap, but neither a large spin gap  nor magnetic ordering is observed down to the lowest temperatures studied (2.1 K) \cite{llusar2004}.
Thus, like other crystals based on dmit complexes it could be a quantum spin--liquid \cite{powell11}.  We show that the  2/3-filled Hubbard model on the three--legged ladder has a non-trivial symmetry protected topological (SPT) phase (the Haldane phase \cite{haldane1983PRL}) that is  stabilized by reflection through the plane perpendicular to the c-axis, cf. Fig. 1, (but not time reversal or dihedral symmetries) even under strong charge fluctuations that suppress the effective spin-one moments. Thus, our calculations also explain the absence of long--range order in \Mo. Understanding SPT phases is important as they violate another central paradigm of condensed matter physics -- that phase transitions occur due to spontaneous symmetry breaking.

\begin{figure}
	\centering
	\includegraphics[width=0.95\columnwidth]{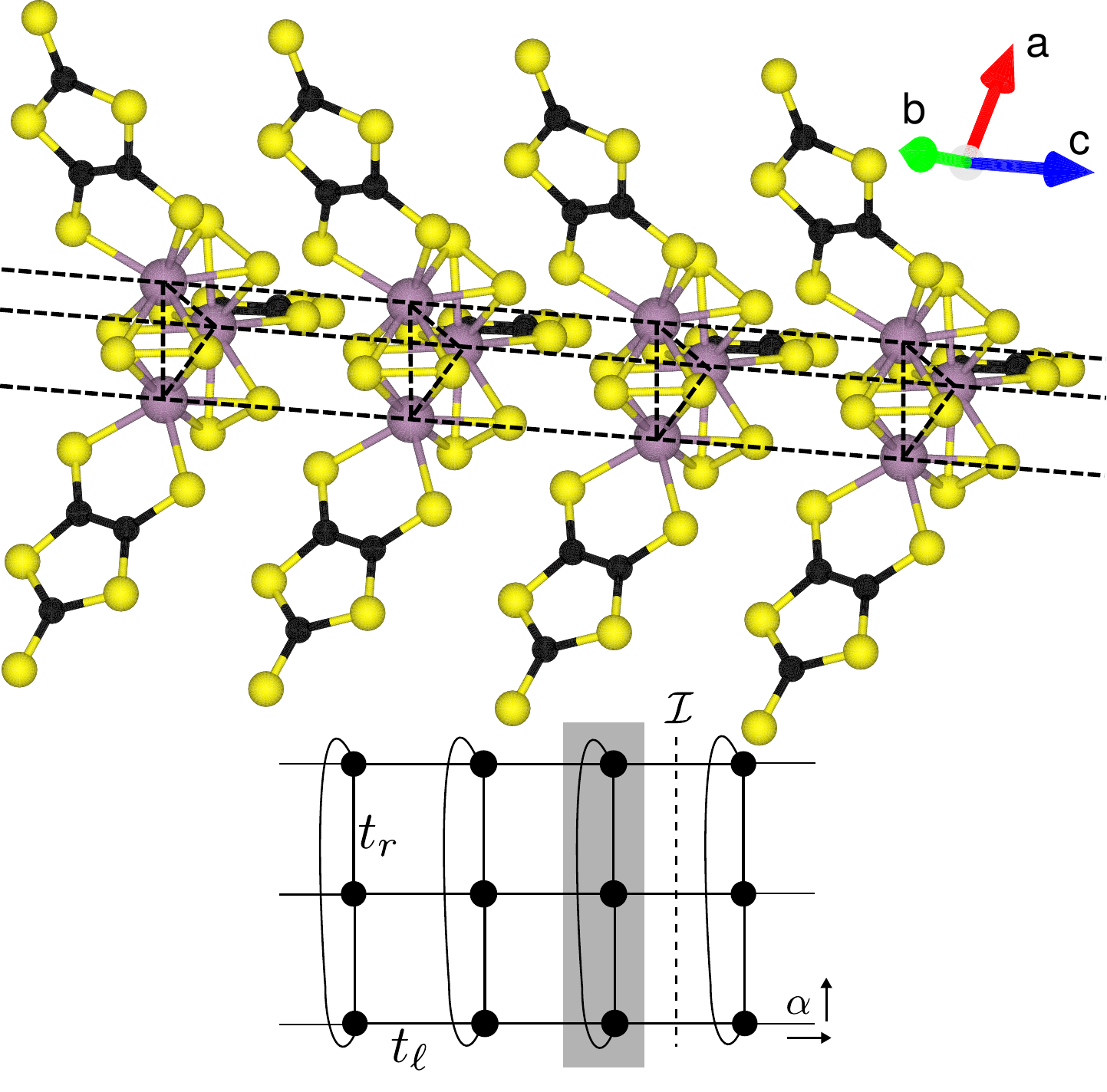}
	\caption{(color online)
		Three-legged ladder in \Mo. The electronic structure of a single molecule is described by three Wannier orbitals that have mostly Mo-dmit hybrid character. \cite{jackoPRB2015} The hopping integral between the Wannier orbitals on a single molecule are denoted $t_r$, while the hopping between equivalent Wanniers on different molecules displaced along the $c$ axis is denoted $t_l$. Interference effects suppress interchain interactions in both the non-interacting \cite{jackoPRB2015} and strongly interacting \cite{merino2016} limits. 
		This results in a three-legged ladder with hopping parameters $t_r$ ($t_{\ell}$) along the rungs (legs) and periodic boundary conditions in the rungs. Reflection about the bond between two triangular molecules is indicated by $\mathcal{I}$. The shaded region indicates the three site unit cell. Based on structural data from. \cite{llusar2004}
		Crystal structure visualized with VESTA. \cite{vesta3}
		\label{fig:model}
	}
\end{figure}

Early band structure calculations for \Mo, which counter-factually assumed antiferromagnetic ordering not seen experimentally, lead to the suggestion that this material is described by models on the `triangular necklace' lattice, with triangular molecules decorated along a one--dimensional (1D) backbone \cite{llusar2004}. The two-thirds filled Hubbard model on the triangular necklace is in the Haldane phase \cite{jananiPRL2014}. However, the formation of spin-one moments in this model depends crucially on a `local parity' symmetry of the model, that is neither generic nor present in the material \cite{jananiPRB2014}. Furthermore, analysis \cite{merino2016} of the Hubbard model with the tight-binding parameters based on Wannier orbitals constructed from paramagnetic density functional calculations \cite{jackoPRB2015} demonstrated that the two-thirds filled three-legged ladder is the relevant Hubbbard model for Mo$_3$S$_7$(dmit)$_3$, cf. Fig. \ref{fig:model}. 

The Hamiltonian of Hubbard model on the three-legged ladder  is 
\begin{align} \label{eq:hamiltonian}
\hat{H} = 
& -t_{\ell}\sum_{i\alpha\sigma}(\hat{c}_{i\alpha\sigma}^{\dagger}\hat{c}_{(i+1)\alpha\sigma}^{} + \mathrm{H.c})
- t_r\sum_{i,\alpha\neq\beta,\sigma}\hat{c}_{i\alpha\sigma}^{\dagger}\hat{c}_{i\beta\sigma}^{} \nonumber \\
& + U\sum_{i\alpha}\hat{n}_{i\alpha\uparrow}^{} \hat{n}_{i\alpha\downarrow}^{},
\end{align}
where $\hat{c}_{i\alpha\sigma}^{(\dagger)}$ annihilates (creates) an electron with spin $\sigma$ on the $\alpha$th leg of the $i$th rung, $\hat{n}_{i\alpha\sigma} = \hat{c}_{i\alpha\sigma}^{\dagger}\hat{c}_{i\alpha\sigma}^{}$, and $t_r$ ($t_{\ell}$) are the hopping amplitudes along the rungs (legs), see Fig.~\ref{fig:model}. 
We employ the infinite density matrix renormalization group (iDMRG) \cite{white1992,mccullocharXiv2008,schollwock2011} method with a matrix product state (MPS) ansatz with SU(2) symmetry ($[\hat{H},\hat{\bm{S}}]=\bm{0}$) \cite{mcculloch2007} keeping $m=2000$ [equivalent to $m \lesssim 6000$ with only U(1) symmetry] basis states. We always use finite basis set scaling to $m\rightarrow \infty$ by using the correlation length for the appropriate symmetry sector. In all figures the error is determined from finite basis set scaling and curves are guides to the eye.

\section{Limiting behaviors}\label{sect:limits}

There are several limits where analytical progress can be made straightforwardly.  For decoupled legs ($t_r=0$) with $t_{\ell}>0$ and $U\rightarrow \infty$ we expect three decoupled chains that are strongly correlated metals; in contrast to half-filling ($n=3$) where a Mott insulating state is formed.  For $t_r, t_{\ell} > 0$ and two-thirds filling ($n=4$ electrons per unit cell) the system is a  metal when $U=0$. For any $U>0$ and in the strong rung limit ($t_{\ell}=0$) the rungs  form triplets ($S=1$) \cite{merino06}. 

Jacko  {\it et al.}\cite{jackoPRB2015} have recently reported density functional calculations for \Mo and constructed a tight-binding model from the Wannier orbitals extracted from the calculation. They reported that $t_{\ell}/t_r\simeq2/3$. However, they also found an interchain hopping $\sim t_{\ell}$. Nevertheless, the single electron band structure is strongly one--dimensional because of interference effects in the in--plane hopping.\cite{jackoPRB2015}

Merino {\it et al.}\cite{merino2016} have recently discussed strongly correlated trinuclear complexes. In the large $U$ limit our model [Eq. (\ref{eq:hamiltonian})] is equivalent to theirs if spin-orbit coupling is neglected. Thus, some insights into the strong--coupling (large $U/t_{\ell}$) molecular (small $t_r/t_{\ell}$) limit of the current model follow immediately from their results.

(1) Effective spin-one moments form on each triangular molecule.

(2) In contrast to the atomic case, the effective nearest neighbor superexchange interaction along the chains, $J_\|$ does not vanish as $U\rightarrow\infty$, rather $J_\|\rightarrow{2t_z^2}/{9t_c}$. This is because neighboring molecules are coupled by three legs, which allows processes contributing to the exchange interaction that do not incur an energy penalty $\sim U$.

(3) In the plane perpendicular to the chains electrons can only hop between a single Wannier orbital on each molecule\cite{jackoPRB2015} (single vertex on each triangle). This means that in the interchain superexchange, $J_\perp\rightarrow0$ as $U\rightarrow\infty$.

(4) Charge fluctuations are significant along the chain, but are very strongly suppressed in the plane.

These results suggest that, at the moderate or large $U$'s expected in the material, a Hubbard model on a three legged ladder is the appropriate model of \Mo. In particular, they suggest that a fermionic description, allowing for the treatment of charge and spin fluctuations on an equal footing, is vital. This adds to the intrinsic interest in ladders.\cite{dagotto1996,deshpande2010}

\section{Numerical results}

\begin{figure}
 \centering
 \includegraphics[width=0.95\columnwidth]{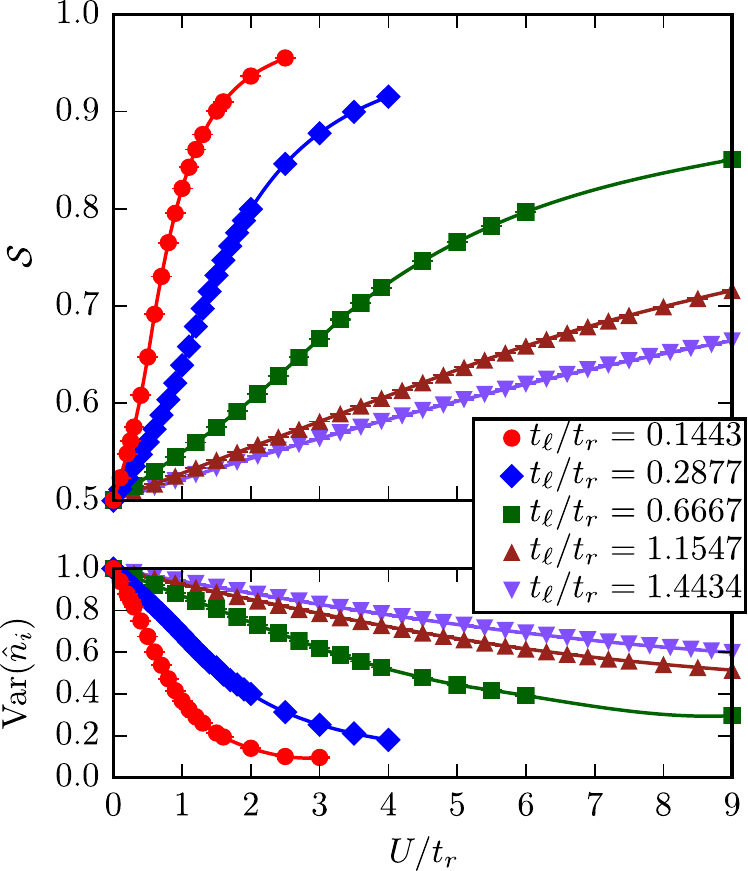}
 \caption{(color online)
	Charge fluctuations [$\mathrm{Var}(\hat{n}_i)$; panel b] cause the effective spin per unit cell [$\mathcal{S}$; panel a] to be suppressed from the ($\mathcal{S}=1$) triplet that is the ground state of an isolated three site cluster. $\mathcal{S}$ is given by the solution of  $\mathcal{S}(\mathcal{S}+1) = \langle\hat{\bm{S}}_i \cdot \hat{\bm{S}}_i \rangle$, where $\hat{\bm{S}}_i^{} = \sum_{\alpha} \hat{\bm{S}}_{i\alpha}^{}$ is the net spin of the $i$th  rung, $\hat{\bm{S}}_{i\alpha}^{} = \sum_{\sigma\sigma'} \hat{c}_{i\alpha\sigma}^{\dagger} \bm{\tau}_{\sigma\sigma'}^{} \hat{c}_{i\alpha\sigma'}^{}$, and $\bm{\tau}$ is the vector of Pauli matrices. 
	$\mathrm{Var}(\hat{n}_i)=\langle\hat{n}_i^2\rangle - \langle\hat{n}_i\rangle^2$, where $\hat{n}_i = \sum_{\alpha\sigma}\hat{n}_{i\alpha\sigma}$. 
	\label{fig:effectivespin}
}
\end{figure}

We investigated parameters ranging from the strongly coupled legs (small $t_{\ell}/t_r$) to  weakly coupled legs (large $t_{\ell}/t_r$) for finite $U$ numerically. We find that increasing $t_\ell$ induces strong on-rung charge fluctuations, Fig.~\ref{fig:effectivespin}b. Nevertheless, significant local moments remain within each unit cell, Fig.~\ref{fig:effectivespin}a, but triplet formation is clearly suppressed as the charge fluctuations increase. 

\begin{figure}
	\centering
	\includegraphics[width=0.95\columnwidth]{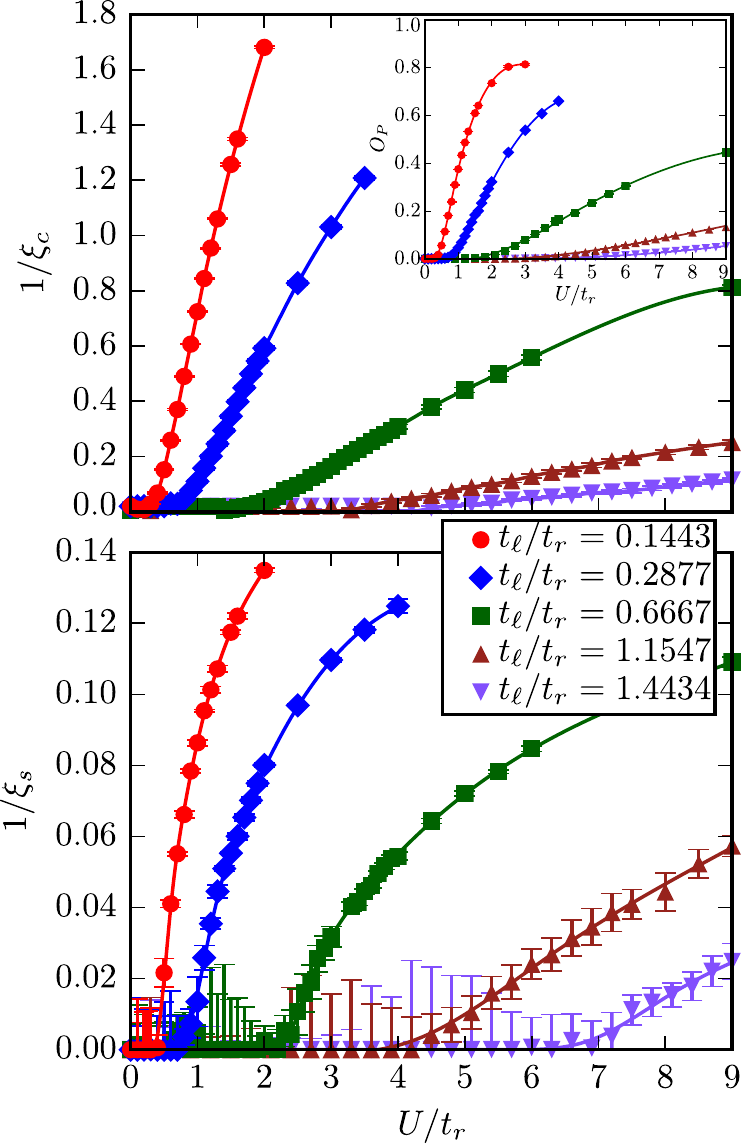}
	\caption{(color online) 
Correlation lengths for the charge ($1/\xi_c$) and spin ($1/\xi_s$) excitations calculated from the transfer matrix of the MPS. $1/\xi_{c}$ is proportional to the charge gap ($\Delta_c$) while $1/\xi_{s}$ is proportional to the spin gap ($\Delta_s$). Inset: Charge string order parameter. $O_P = \lim_{|i-j|\rightarrow\infty} \langle \mathbb{1}_i \exp(i\pi\sum_{l=i}^{j} \hat{n}_l ) \mathbb{1}_j \rangle$, where $\mathbb{1}_i^{}$ is the identity operator on the $i$th set of rungs and $\hat{n}_l = \sum_{\sigma} \hat{n}_{l\sigma}$.  $O_P=0$ in the metallic phase, but $O_P\neq 0$ for a charge ordered state indicating that molecular (on-rung) chargons are bound in the insulating phase. 
		\label{fig:gaps}
	}
\end{figure}

We find clear evidence of an insulating ground state. 
The correlation length, $\xi$, is determined from the transfer matrix of the MPS. Area laws imply that a spectral gap in a 1D system corresponds to a finite correlation length \cite{eisertPRev2010}, such that $1/\xi_{c}$ ($1/\xi_{s}$) is proportional to the charge (spin) gap. 
We find both gaps remain finite, although become numerically non-trivial to distinguish from zero for small $U/t_r$ when scaled $m\rightarrow \infty$, Fig.~\ref{fig:gaps}.
%
%
The gaps can also be directly calculated using finite DMRG and consistent results are found, and, e.g., we find that $\Delta_c/t_r\sim 0.8$ and $\Delta_s/t_r\sim 0.007$ for $t_{\ell}/t_r=0.1443$ at $U/t_r=2$. With finite DMRG we also find at large $U$ that $\Delta_c\sim U$ and $\Delta_s$ saturates, in agreement with behavior found in $1/\xi_{c}$ and $1/\xi_{s}$ from iDMRG. Note that both methods give a charge gap that is orders of magnitude larger than the spin gap. 

The physics for small $U/t_r$ remains somewhat ambiguous. The gap certainly becomes much smaller and may close. However, in the ambiguous region the correlation length does not saturate for any $m$ and the basis set scaling becomes extremely difficult. It is possible that in this region for some parameters there is a transition to a metal that we are unable to observe. This ambiguity is consistent with a gapless ground state  close in energy to the SPT state because DMRG is biased in favor of gapped states due to the finite basis.\cite{schollwock2011} Because of these uncertainties we confine the discussion below to the unambiguous region where the gaps are clearly finite.

At half-filling the Mott metal-insulator transition occurs because of the binding of chargon (doublon or holon) pairs \cite{mott1949,gebhard1997}.
The low-energy charge excitations of, say, the 1D Bose-Hubbard model are highly analogous to the three spin states in the Heisenberg model. Thus a  charge string order parameter  can be defined for the superfluid-insulator transition \cite{dallatorrePRL2006}. The binding of chargons is attested by a dramatic increase in the charge string order parameter  in the insulating phase \cite{Endres}. 

A natural explanation for the insulating phase at two-thirds filling in the triangular ladder emerges if one views each unit cell as a triangular molecule, as is appropriate in \Mo. \cite{jackoPRB2015,merino2016} There is a marked tendency to form triplets on each molecule (rung), cf. Fig. \ref{fig:effectivespin} and Refs. \onlinecite{jananiPRB2014,jananiPRL2014}. This may tend to bind molecular chargons (i.e., a local excess of charge once the rung degree of freedom is integrated out). This hypothesis can be tested by  calculating a rung-charge string order parameter:
\begin{equation}
O_P = \lim_{|i-j|\rightarrow\infty} \langle \mathbb{1}_i \exp(i\pi\sum_{l=i}^{j} \hat{n}_l ) \mathbb{1}_j \rangle,
\end{equation}
where $\hat{n}_l = \sum_{\sigma} \hat{n}_{l\sigma}$. If the insulating phase  results from the binding of molecular chargons one will find that $O_P\neq 0$ in the insulating phase and $O_P \rightarrow 0$ in the metallic phase.  We find that whenever there is unambiguously a (charge) gapped ground state $O_P$ is large,  Fig.~\ref{fig:gaps} inset.  $O_P$ is numerically zero at $U=0$, as expected \cite{chargestring}, and extremely small where the finite basis set scaling is non-trivial. In the strong rung limit and $U\rightarrow\infty$, we find $O_P$ approaches unity; away from this limit $O_P$ is suppressed, suggesting that the molecular chargons become progressively more weakly bound. Hence, we conclude that at finite $U$ the model \eqref{eq:hamiltonian} is a correlated insulator that is not of the usual Mott type, where on-site interactions bind atomic chargons. At two-thirds filling  on-rung triplet formation binds molecular chargons. 

In the strong rung and large $U$ limits with the formation of triplets on the rungs, superexchange 
between different sets of rungs would suggest that an effective low-energy Hamiltonian is the spin-one Heisenberg chain \cite{merino2016} with the Haldane phase as its ground state \cite{haldane1983PRL}. 
We will demonstrate below that  model \eqref{eq:hamiltonian} remains in the Haldane phase throughout the insulating phase -- even away from these limits, regardless of the suppression from unity of the spin on the rungs (${\cal S}\ll 1$) seen in Fig.~\ref{fig:effectivespin}.

\begin{figure}
	\centering
	\includegraphics[width=0.95\columnwidth]{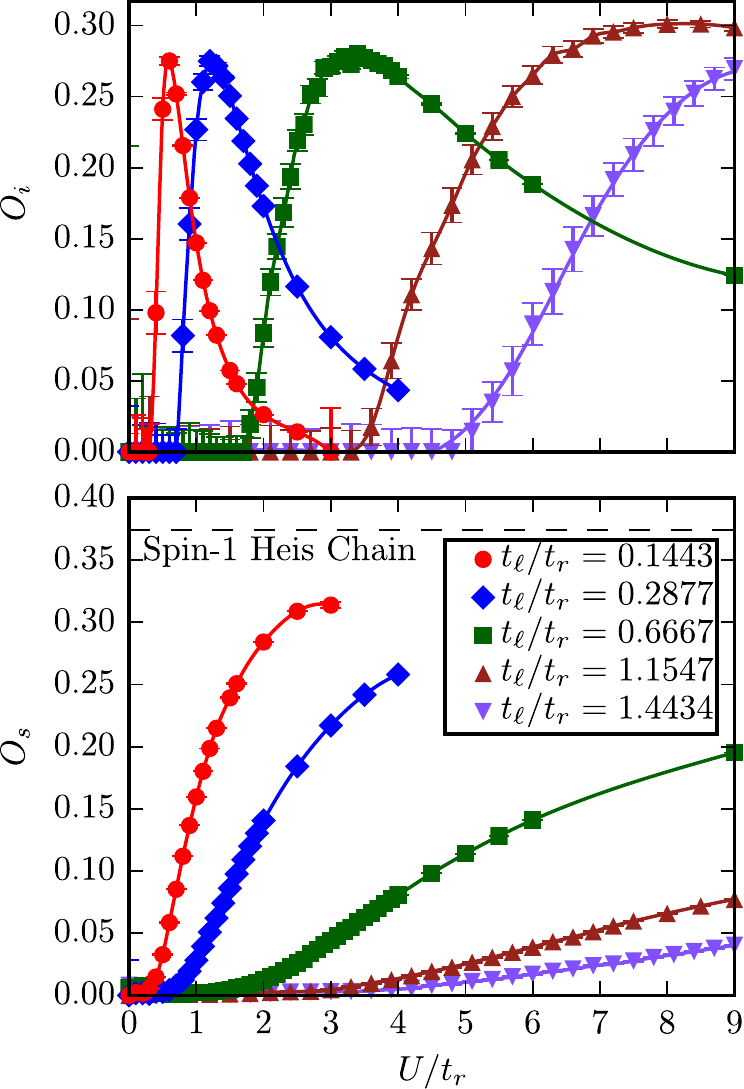}
	\caption{(color online) 
		String order parameters $O_i=\lim_{|i-j|\rightarrow \infty} \langle \mathbb{1}_i^{} \exp(i\pi \sum_{l=i+1}^{j-1} \hat{S}_l^z) \mathbb{1}_j^{} \rangle$ and $O_s= - \lim_{|i-j|\rightarrow \infty} \langle \hat{S}_i^z \exp(i\pi \sum_{l=i+1}^{j-1} \hat{S}_l^z) \hat{S}_j^z \rangle$.
		As expected $O_i$ and $O_s$  disappear  as the spin gap, $\Delta_s$, closes. As $U/t_{\ell}\rightarrow\infty$ and $t_{\ell}/t_r \rightarrow \infty$ the string order parameters  approach values found in the ground state of the spin-one Heisenberg chain, where $O_i=0$ \cite{pollmannPRB2012} and $O_s = 0.374\,325\,096(2)$ \cite{whitePRB1993}.
		\label{fig:spinorder}
	}
\end{figure}

The $D_2$ symmetry of the spin-one Heisenberg model naturally leads to the study of a pair of string order parameters\cite{nijs1989,tasaki1992,pollmannPRB2012}
\begin{equation}
O_i=\lim_{|i-j|\rightarrow \infty} \langle \mathbb{1}_i^{} \exp(i\pi \sum_{l=i+1}^{j-1} \hat{S}_l^z) \mathbb{1}_j^{} \rangle,
\end{equation}
and 
\begin{equation}
O_s= - \lim_{|i-j|\rightarrow \infty} \langle \hat{S}_i^z \exp(i\pi \sum_{l=i+1}^{j-1} \hat{S}_l^z) \hat{S}_j^z \rangle,
\end{equation}
where $\hat{S}_i^{z} = \sum_{\alpha} \hat{S}_{i\alpha}^{z}$ is the spin  of the $i$th unit cell projected onto the $\hat{z}$ axis, $\hat{S}_{i\alpha}^{z} = \sum_{\sigma\sigma'} \hat{c}_{i\alpha\sigma}^{\dagger} \tau_{\sigma\sigma'}^{z} \hat{c}_{i\alpha\sigma'}^{}$, $\tau_{\sigma\sigma'}^{z}$ is the Pauli matrix, and $\mathbb{1}_i^{}$ is the identity on the $i$th unit cell. These string order parameters are plotted in Fig.~\ref{fig:spinorder}.
Independently neither identifies a topological phase 
\cite{ciracPRL2008, pollmannPRB2012}. Instead, there are selection rules for spin models \cite{pollmannPRB2012}, whereby $O_i=0 \,\, (\neq 0)$ and $O_s\neq 0 \,\, (=0)$ in the Haldane (trivial) phase.
In the present fermionic model,  both $O_i$ and $O_s$ remain finite, and hence these selection rules cannot determine whether there is an SPT phase, Fig.~\ref{fig:spinorder}. 
Note also that both $O_i$ and $O_s$ vanish  when the spin gap closes, much as $O_P\rightarrow0$ when  the charge gap closes.  

\begin{figure}
 \begin{center}
 \includegraphics[width=0.95\columnwidth]{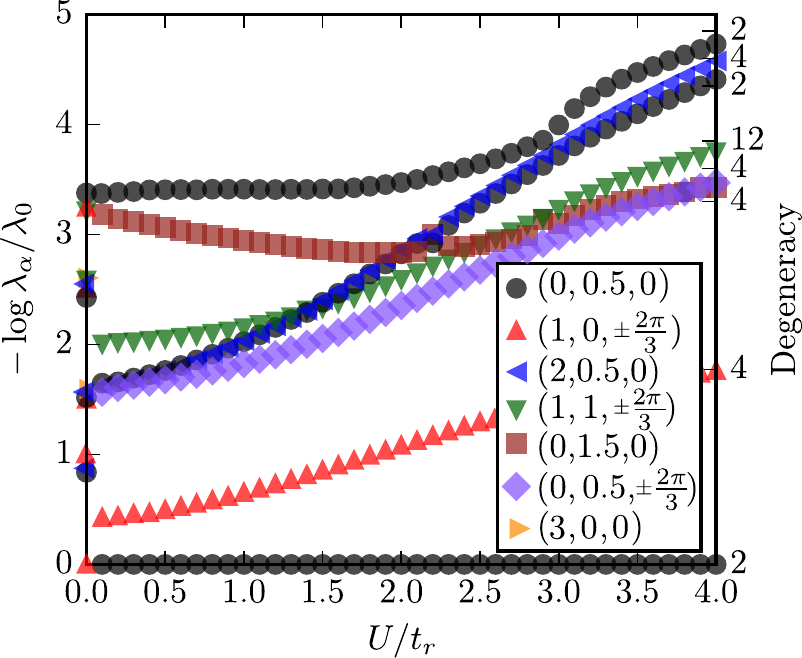}
\end{center}
 \caption{(color online) 
The low-lying eigenvalues $\lambda_{\alpha}$ of the entanglement spectrum, i.e., the eigenvalues of the reduced density matrix on tracing out half of the system, for $t_{\ell}/t_r = 0.6667$. Here (N,S,k) label particle, spin, and momentum around a triangular molecule quantum numbers respectively. The entanglement spectrum remains even-fold degenerate for all non-zero $U/t_r$; degeneracies are indicated by the numbers on the right axis. At $U=0$ odd-fold degeneracies are also found consistent with the expected trivial metal. The even-fold degeneracy is a robust signature of an SPT state.
\label{fig:entanglement}
}
\end{figure}

A robust signature of a non-trivial SPT state is an even-fold degeneracy in the entanglement spectrum \cite{pollmannPRB2010}, i.e., the eigenvalues of the reduced density matrix on tracing out half of the system. For the Haldane phase the even-fold degeneracy in the entanglement spectrum is a result of degenerate spin-1/2 edge states. We find that the entanglement spectrum of the current model remains even-fold degenerate for any finite $U$, Fig.~\ref{fig:entanglement},  indicating a non-trivial SPT phase. It is interesting to note that this is true even in the region where existence of the spin gap is ambiguous. However, finite basis set scaling is of limited efficacy for the entanglement spectrum and so this result should not be over interpreted. The entanglement spectrum includes odd-fold degeneracies at $U=0$ consistent with the known topologically trivial metal.

However, the entanglement spectrum  does not uniquely specify the SPT phase \cite{pollmannPRB2010}. One can characterize an SPT phase by the projective representation of its symmetry groups \cite{chen2011a,schuch2011,pollmannPRB2012}. Since the state is invariant under these internal symmetries  
\begin{equation}
\sum_{jj'}R_{jj'}^{(g)}A_j = e^{i\theta_g}U_g^{\dagger}A_jU_g,
\end{equation}
where $R_{jj'}^{(g)}$ is a unitary matrix that transforms the MPS matrices, $A_j$, under the relevant symmetry operation, $g$, $U_g$ is a unitary matrix, and $e^{i\theta_g}$ is a phase factor. The unitary matrices  form a projective representation of the symmetry group. Since it is not possible to continuously deform a state between two different SPT phases without going through a phase transition, $\theta_g$ classifies the different topological phases. In spin models the  Haldane phase has a non-trivial projective representation with $\theta_g=\pi$ for all protecting symmetries ($Z_2\times Z_2$, TR, $\mathcal{I}$). Note that in the limit of a spin chain inversion and the reflection about the bond between two triangular molecules are equivalent. If all the phase factors are identities, the representation is linear; this describes a topologically trivial phase, which can be adiabatically connected to a product state while preserving the protecting symmetries.

For an infinite MPS we can directly calculate $\theta_g$ from the `non-local' order parameters;\cite{haegeman2012,pollmannPRB2012} for $Z_2\times Z_2$
\begin{equation}
O_{Z_2\times Z_2}=\langle U_y^{} U_z^{} U_y^\dagger U_z^\dagger \rangle,
\end{equation}
where $U_\mu$  is a rotation about the $\mu$ spin axis; for time-reversal 
\begin{equation}
O_{\mathrm{TR}} = \langle U_{\mathrm{TR}}^{} U_{\mathrm{TR}}^* \rangle,
\end{equation}
and for reflection, 
\begin{equation}
O_{\mathcal{I}} = \langle U_{\mathcal{I}}^{} U_{\mathcal{I}}^* \rangle.
\end{equation}
For spin-one chains $O_g=1$ in the trivial phase while $O_g=-1$ in the Haldane phase. If the symmetry does not protect the state then $|O_g|<1$.

\begin{figure}
	\centering
	\includegraphics[width=0.95\columnwidth]{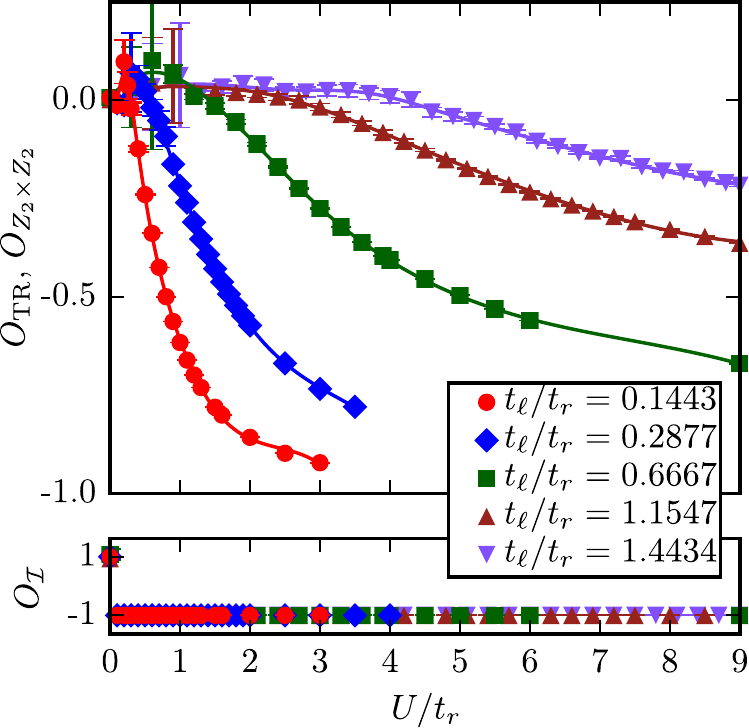}
	\caption{(color online)
		Order parameters $O_{Z_2\times Z_2}= \langle U_y^{} U_z^{} U_y^{\dagger} U_z^{\dagger} \rangle$, $O_{\mathrm{TR}} = \langle U_{\mathrm{TR}}^{} U_{\mathrm{TR}}^* \rangle$, and $O_{\mathcal{I}} = \langle U_{\mathcal{I}}^{} U_{\mathcal{I}}^*\rangle$, where $U$ is a unitary matrix resulting from transforming the iMPS state by dihedral ($D_2\cong Z_2\times Z_2$), time-reversal (TR), or reflection ($\mathcal{I}$) symmetries respectively. Note that $O_{\mathrm{TR}}=O_{Z_2\times Z_2}$ because the time-reversal transformation is $\mathcal{K}\exp(-i\pi \hat{S}_y)$ where $\mathcal{K}$ is complex conjugation. $O_g=1$ in the trivial  phase and $O_g=-1$ in the non-trivial SPT phase, whereas $|O_g| < 1$ if the symmetry is not protecting. Only reflection symmetry stabilizes the non-trivial SPT phase (Haldane phase) for finite $U/t_r$, even under strong charge fluctuations.
		\label{fig:symmreps}
	}
\end{figure}

We find $Z_2\times Z_2$ and time reversal symmetries are not protecting in this model due to charge fluctuations, see Fig.~\ref{fig:symmreps}, consistent with previous results for other fermionic models  \cite{anfuso2007PRB, moudgalya2015PRB}. This conclusion is supported by the mixing of half-integer and integer spin excitations observed in the entanglement spectrum. Half-integer spin excitations are in the non-trivial projective representation of $Z_2\times Z_2$ and are antisymmetric under time reversal, whereas integer spin excitations are in the linear representation and time-reversal symmetric. At $U=0$ there is an even mix of half-integer and integer spin excitations, indicated by $O_{Z_2\times Z_2} = O_{\mathrm{TR}} = 0$. As $U/t_r\rightarrow \infty$ and $t_\ell/t_r\rightarrow 0$, we asymptotically recover  the non-trivial projective representation of half-integer spin excitations in the entanglement spectrum as charge fluctuations are suppressed, and we arrive at the spin-one Heisenberg chain with the Haldane phase as its ground state. This is further indicated in the string order parameter, Fig.~\ref{fig:spinorder}, where the selection rules are recovered: $O_i\rightarrow0$ and $O_s\neq 0$ in this limit.

On the other hand, reflection symmetry is always protecting. $O_{\mathcal{I}} = -1$ when scaled  $m\rightarrow \infty$ for any   $U \neq 0$, indicating the model is in the Haldane phase, see Fig.~\ref{fig:symmreps}. At $U=0$, there is a jump to $O_{\mathcal{I}} =1$ as expected of a phase transition to the trivial metal. 
Clearly as real materials are composed of fermions the Haldane phase is only protected by reflection symmetries, with $Z_2\times Z_2$ and time-reversal additionally protecting only in approximate spin models.


\section{Conclusions}

It is natural to ask whether similar insulating phases occur for other ladders at commensurate fillings? The molecular perspective suggests that this should be the case when Nagaoka's theorem \cite{nagaoka1966} holds for an isolated unit cell, e.g., $(\ell+1)/2\ell$-filling for the $\ell$-leg ladder with $t_\ell>0$. For even $\ell$ this would imply half-odd-integer spin and a vanishing spin gap. For $\ell>3$ there may also be an intermediate $U$ insulating phase when Hund's rules dominate. \cite{jananiPRB2014}

While the model is interesting in its own right, it is also important to ask how robust the connection to materials, such as \Mo, is. For example, is there a phase transition back to a metallic state for weakly coupled legs (large, but finite, $t_{\ell}/t_r$)? What role does inter-ladder coupling play in real materials? Analysis of the strong coupling molecular limit\cite{merino2016} (see, also, Sec.~\ref{sect:limits}) suggest that for the intermediate-to-large $U$ one expects in \Mo interchain coupling is strongly suppressed and so the above results in the region with large spin and gaps should be robust. However, for small enough $U$ the inter-ladder effects will become non-negligible. This suggests that, even if the gap remains non-zero but small in the region where our numerics are ambiguous, it may well be destroyed by inter-ladder coupling. Thus a phase transition from a metal to a correlated insulator as a function of $U/t_\ell$ would not be surprising. \Mo is insulating, suggesting it is in the large $U/t_\ell$ regime. In other correlated molecular crystals the application of hydrostatic pressure is believed to decrease $U$, relative to the intramolecular hopping. This suggests that in \Mo the application of pressure (or uniaxial strain) could drive the system metallic.

Other important questions asked by our work include: Do the molecular (rung) triplets drive superconductivity if the full 3D model is doped away from two-thirds filling or on the applications of hydrostatic pressure? If so how analogous is this to the resonating valence bond theory \cite{Anderson87,Powell05,Powell07} of superconductivity near the Mott insulator? And, is this superconductivity realized in A$_2$Cr$_3$As$_3$ (A=K, Rb, Cs) \cite{wuNatComms2014} or LiMoO  \cite{greenblatt1984,merinoPRB2012}?

\begin{acknowledgments}
We thank Seyed Nariman Saadatmand for helpful conversations. This work was supported by the Australian Research Council through grants FT130100161, DP160100060, CE110001013, FT140100625 and LE120100181. Calculations were performed with resources from the National Computational Infrastructure (NCI), which is supported by the Australian Government, and with the compute resources of Obelix in the School of Mathematics and Physics at the University of Queensland.
\end{acknowledgments}


\begin{thebibliography}{44}
\bibitem [{cha()}]{chargestring}%
\BibitemOpen
\href@noop {} {}\bibinfo {note} {{N}ote that for a three-site cluster there
	are equals numbers of configurations with odd and even numbers of electrons:
	$\sum_{n=0}^3 {6 \choose 2n} = \sum_{n=0}^2 {6 \choose 2n+1} = 32$. In a
	correlated metal one may find small non-zero values of $O_P$ because the
	model is not particle-hole symmetric.}\BibitemShut {Stop}%
\end{thebibliography}

\end{document}